\documentclass[conference]{IEEEtran}
\IEEEoverridecommandlockouts
\usepackage{comment}
\usepackage{amsmath,amsfonts}
\usepackage{array}
\usepackage[caption=false,font=normalsize,labelfont=sf,textfont=sf]{subfig}
\usepackage{textcomp}
\usepackage{stfloats}
\usepackage{url}
\usepackage{verbatim}
\usepackage{graphicx}
\usepackage{array}
\usepackage{caption} 
\usepackage{xcolor}
\usepackage{multirow}
\usepackage[ruled, lined, linesnumbered, longend]{algorithm2e}
\usepackage{soul} 

\def\BibTeX{{\rm B\kern-.05em{\sc i\kern-.025em b}\kern-.08em
    T\kern-.1667em\lower.7ex\hbox{E}\kern-.125emX}}
    
\SetKwInput{KwOutput}{Output}   
\SetKwInput{KwInput}{Input}

\newcommand{\fixme}[1]{{\textcolor{red}{\em\bf{[FIXME: #1]}}}}

\begin{document}

\title{\textit{Trauma lurking in the shadows}: A Reddit case study of mental health issues in online  posts about Childhood Sexual Abuse}

\author{
\IEEEauthorblockN{Orchid Chetia Phukan}
\IEEEauthorblockA{\textit{Department of CSE} \\
IIIT-Delhi, India\\
orchidp@iiitd.ac.in}`
\and
\IEEEauthorblockN{Rajesh Sharma}
\IEEEauthorblockA{\textit{Institute of Computer Science} \\
University of Tartu, Estonia\\
rajesh.sharma@ut.ee}
\and
\IEEEauthorblockN{Arun Balaji Buduru}
\IEEEauthorblockA{\textit{Department of CSE} \\
IIIT-Delhi, India\\
arunb@iiitd.ac.in}
}

\maketitle

\begin{abstract}
Childhood Sexual Abuse (CSA) is a menace to society and has long-lasting effects on the mental health of the survivors. From time to time CSA survivors are haunted by various mental health issues in their lifetime. Proper care and attention towards CSA survivors facing mental health issues can drastically improve the mental health conditions of CSA survivors. Previous works leveraging online social media (OSM) data for understanding mental health issues haven't focused on mental health issues in individuals with CSA background. Our work fills this gap by studying Reddit posts related to CSA to understand their mental health issues. Mental health issues such as depression, anxiety, and Post-Traumatic Stress Disorder (PTSD) are most commonly observed in posts with CSA background. Observable differences exist between posts related to mental health issues with and without CSA background. Keeping this difference in mind, for identifying mental health issues in posts with CSA exposure we develop a two-stage framework. The first stage involves classifying posts with and without CSA background and the second stage involves recognizing mental health issues in posts that are classified as belonging to CSA background. The top model in the first stage is able to achieve accuracy and f1-score (macro) of 96.26\% and 96.24\%. and in the second stage, the top model reports hamming score of 67.09\%.

\textbf{Content Warning: Reader discretion is recommended as our study tackles topics such as child sexual abuse, molestation, etc.}
\end{abstract}



\section{Introduction}
Sexually harmful acts against a child (less than 18 years of age) are characterized as Childhood Sexual Abuse (CSA) and it spans a wide range of acts including sexual abuse, sexual interaction, rape, sexual grooming, and sexual exploitation \cite{murray2014child}. 
This form of abuse often entails the perpetrator using force or making threats \cite{browne1986impact}. Between ages 7 and 13, children are most vulnerable to CSA. Perpetrators can be either family members or strangers \cite{Finkelhor1994-qn}. It has been reported that survivors of CSA are often accompanied by feelings of stigma and guilt \cite{coffey1996mediators, kennedy2018still}.\par

CSA is often succeeded by physical and mental difficulties. Physical difficulties include physical injuries such as genital injuries \cite{mccann1992genital} and sexually transmitted diseases (STDs) \cite{johnson2004child}. Individuals with a history of CSA are more likely to have chronic illnesses such as chronic pain \cite{kamiya2016impact} and obesity \cite{gustafson2004childhood} in later stages of life. Physical injuries may heal, but psychological scars from the abuse can last throughout later stages of life \cite{johnson2004child}. Survivors of CSA are often likely to be haunted by the recall of the traumatic event \cite{loftus1994memories}. Risk of occurrence of various mental health issues such as  Post-Traumatic Stress
Disorder (PTSD) \cite{briere1987post}, depression, anxiety \cite{kamiya2016impact}, and eating disorders \cite{wonderlich1997relationship} are more in CSA survivors. CSA is also reported to be a key risk factor for the development of Borderline Personality Disorder (BPD) \cite{de2018borderline}.\par

For the purpose of understanding mental health issues in CSA survivors, numerous prior research investigations have been conducted, but they were primarily restricted to interviews, questionnaires, surveys, and electronic health records (EHRs) with a limited amount of data \cite{banyard2001long, werbeloff2021childhood}. In recent years research considering OSM for mental health issues is rising \cite{chancellor2020methods}, as it has been seen that individuals are resorting to online social media (OSM) platforms to find shelter for their mental health issues \cite{bucci2019digital} where CSA survivors are no exception. Especially OSM such as Reddit has turned out to be a popular medium \cite{fraga2018online} for such. Reddit is a community-based platform where each community is called a ``subreddit'' and is based on a particular topic or issue \cite{de2014mental}. These communities sometimes act as a discussion forum and sometimes as peer support groups for various issues related to health, mental complications, etc. Thus, this inspired us to look through various subreddits and gather posts about mental health issues associated with CSA, and probe the following three research questions.


\textbf{RQ1: What are the commonly discussed mental health issues in posts related with CSA?} \par

For finding out the most commonly discussed mental health issues in posts with CSA exposure, we analyzed \textit{r/adultsurvivors}. We found out that depression, anxiety and PTSD are the most commonly discussed mental health issues. 

\textbf{RQ2: Are there differences between the posts related to mental health issues with and without CSA background?} \par
Mental health issues can be triggered by reasons other than CSA. So, in order to identify the specific characteristics of mental health issues related posts with CSA background which could be different from posts without CSA background, we make use of NLP techniques such as word-shift, word cloud, topic analysis, and emotion analysis. We found that posts with CSA background mention flashbacks of the traumatic event. Trauma-related issues are more prevalent in posts with CSA background. Emotion ``disgust'' and ``fear'' are more associated with posts with CSA background in contrast to posts without a CSA background which are more associated with emotion ``sadness''.\par

\textbf{RQ3: Can fine-tuning domain-specific transformer model help in identifying mental health issues in CSA survivors from the OSM posts?} \par

The analysis in RQ2 revealed that indeed there are differences between mental health issues-related posts with and without CSA background. So we developed a two-stage framework where the first stage is binary classification of posts to ``with CSA'' and ``without CSA'' background. We implemented various classical machine learning models and transformer-based models for it. 
Among all the models, fine-tuned BERT (Bidirectional Encoder Representations from Transformers) \cite{devlin2018bert} reported the highest accuracy of 96.26\% and F1-score (Macro) of 96.24\%. In the second stage, multi-label classification is done on the posts with CSA background for the identification of different mental health issues. Here, we fine-tune MentalBERT \cite{ji2021mentalbert}, which is a mental health domain-specific language model and it reported the highest hamming score of 67.09\%. In addition, we provide interpretations of the model's decision using the Integrated Gradients method for both stages.
\par


\textbf{Contributions:} 
Our work is the first, to the best of our knowledge which has analyzed OSM for the understanding and identification of mental health issues in CSA survivors. We present the commonly discussed mental health issues confronted by CSA survivors and the distinctive characteristics of the posts with CSA background that differ from the posts without CSA background. We created an anonymized dataset of around 8k posts related to CSA collected from various subreddits for carrying out our analysis. We highlight the existence of minor distinctions between the posts with and without CSA background and with consideration of this, we propose a two-stage framework for detecting mental health issues in posts with CSA background. 


\textbf{Ethical and Privacy concerns:} Although our institutional review board waived review of our work, we followed preventive measures to protect the privacy of the users whose data is being used in our study. No sensitive or meta information related to the users is collected or considered during our analysis. The examples of posts shown in our study have been paraphrased to avoid traceability. The predictive model only takes the posts and no user information is required. In addition, the work does not propose any diagnostic claims.

\section{Background}
In this section, we first discuss existing literature related to mental health in CSA survivors in Subsection \ref{rw:csa} and then works that have used OSM to understand mental health issues in Subsection \ref{subs0}.

\subsection{Childhood sexual abuse (CSA) and mental health}\label{rw:csa}

A direct relationship between CSA and severe mental health issues has been observed in previous studies \cite{banyard2001long,nelson2002association}. Studies have discovered that CSA has a strong association with serious mental health issues in women who were victims of CSA. Substance addiction and suicidal behavior were also more prevalent among CSA survivors \cite{mullen1993childhood}. 
Research focused on children exposed to CSA (both male and female, 16 years of age or younger) reported that children exposed to CSA are substantially more likely than the general population to need psychiatric treatment. Schizophrenia was the least common mental health issue among CSA survivors of all the mental health issues \cite{spataro2004impact}. Men exposed to CSA are at higher risk of severe mental health issues and also suicidality  \cite{easton2013suicide}. Electronic health records (EHRs) were also availed for a better understanding of the relationship between CSA and serious mental health issues. It was found that patients who have been exposed to CSA are more likely to develop clinical depression, PTSD, and personality disorders \cite{werbeloff2021childhood}.

\vspace{-0.3cm}

\subsection{Mental health issues and online social media (OSM)} 
\label{subs0}
Various research studies using OSM data \cite{Kim2020,vajre2021psychbert} for analyzing and identifying mental health issues have been conducted during the previous decade and this has proven to be an effective method \cite{chancellor2020methods}. These studies are divided into two groups: those that focus on a single mental health issue and those that focus on multiple mental health issues. Reddit posts were leveraged to identify depressive disorder using multiple predictive modeling approaches such as support vector machine (SVM) and multilayer perceptron (MLP) \cite{tadesse2019detection}. Furthermore, using a multi-modal (text, image, etc) strategy, Chiu et al. \cite{chiu2021multimodal} investigated the identification of depressive disorder. Using tweets from self-proclaimed schizophrenia patients, Mitchell et al. \cite{mitchell2015quantifying} looked into transmitting linguistic markers for schizophrenia detection. Multiple BERT-based algorithms were used to identify eating disorders using tweets \cite{benitez2021bert}. \par

Suicidal Intent detection is a crucial task for early detection and prevention of suicide. For identification of suicidal intent, Facebook posts were used \cite{ophir2020deep}. Using RoBERTa, investigation was done for the identification and classification of several mental health issues (depression, anxiety, bipolar disorder, Attention deficit hyperactivity disorder (ADHD), PTSD) \cite{murarka2020detection}. For achieving it, Reddit posts were exploited. Previous research also looked into using a two-stage approach for detecting mental health issues, with the first stage involving binary classification into mental health or non-mental health, and the second stage involving multi-class classification of mental health issues \cite{gkotsis2017characterisation,vajre2021psychbert}. Recently COVID-19 pandemic took a toll on people's mental health which motivated researchers
to look into different mental health subreddits such as
\textit{r/SuicideWatch}, \textit{r/depression}, etc. to investigate the impact of different mental health issues on individuals during the COVID-19 pandemic's early waves \cite{low2020natural}. Past research studies involving OSM either ignored or addressed mental health concerns in CSA survivors as part of a larger study of mental health issues in the general community, with no specific emphasis placed on it. Our work takes a step in this direction.

\section{RQ1:  What are the commonly discussed mental health issues in posts related with CSA?} \label{sub2}

We strive to comprehend the different mental health issues that CSA survivors face. For achieving it, we explored \textit{r/adultsurvivors}
\cite{Stanton2017-ze} subreddit defined as ``a peer aid community for adults who experienced sexual abuse as children'', which provides a platform for CSA survivors to share about their experiences and seek support for their difficulties. We preferred Reddit over other OSM platforms as it is a community-focused platform and also provides a distinctive feature called ``throwaway'' accounts to its users. This temporary feature acts as a veil and allows the users to anonymously express their feelings or subjects that seem inexpressible \cite{de2014mental}.
\\

\textbf{Collecting posts with CSA background:} For finding out commonly discussed mental health issues among CSA survivors, first we collected posts from \textit{r/adultsurvivors} using keywords related to various mental health issues such as depression (\textit{depression}, \textit{depressed}, \textit{hopelessness}, \textit{hopeless}, \textit{depressive}, and \textit{despondent}), anxiety (\textit{anxiety}, \textit{panic attack}, \textit{panic disorder}, and \textit{phobia}), PTSD (\textit{ptsd}, \textit{post-traumatic stress disorder},  \textit{posttraumatic stess disroder}, \textit{trauma}, \textit{traumatic}, and \textit{traumatized}), schizophrenia (\textit{schizophrenia} and \textit{psychosis}), eating disorders (\textit{eating disorder}, \textit{anorexia}, and \textit{bulimia}), and personality disorders (\textit{bpd}, \textit{personality disorder}, and \textit{ocpd}). We got the highest number of posts related to depression, anxiety, and PTSD. We got few to none posts related to mental health issues such as schizophrenia, eating disorders, and personality disorders. Since depression, anxiety, and PTSD are the most commonly discussed mental health issues so we focus only on these three mental health issues and discarded other mental health issues for carrying out our further analysis. \par

We didn't collect posts from subreddit such as \textit{r/survivorsofabuse}, a support group for abuse survivors \cite{teblunthuis2022identifying}, as it is a private community. We collected posts from the subreddits that are publicly accessible. Taking inspiration from aforementioned, we also collected posts with CSA background from various mental health subreddits associated with a particular mental health issue such as \textit{r/depression}, \textit{r/Anxiety}, \textit{r/ptsd}, and \textit{r/depression\_help} \cite{murarka2020detection} related to commonly discussed mental health issues using keywords such as \textit{childhood sexual abuse} and \textit{csa}. We collected posts related to commonly discussed mental health issues from \textit{r/mentalhealth} (most popular health-related subreddit) \cite{gaur2018let} using the same keywords used for collecting posts in \textit{r/adultsurviors} and also used keywords such as \textit{childhood sexual abuse} and \textit{csa} to make sure they are related to CSA. We collected posts posted by normal accounts as well as throwaway accounts. Posts from the subreddits are collected using PushShift API. Posts from the above-mentioned subreddits are from the date these subreddits started till January 2022. Along with the title and post, author ID's for each post is collected. Multiple posts may have the same author ID. These posts are subjected to a manual inspection to ensure that they are CSA-related. Table \ref{tab:table1} shows information about posts collected through various subreddits. ``\# of posts'' column shows the count of posts collected from each subreddit. We got the highest number of posts from \textit{r/adultsurvivors}.\par

We dropped posts tagged as ``[deleted]'' because these posts are either deleted or posted by users who have deleted their accounts and also posts tagged as ``[removed]'' are dropped as these posts are removed by the moderator of the subreddit or spam filter. Table \ref{tab:table1} ``Final \# of posts'' column presents the final count of posts from each subreddit. Subreddits \textit{r/depression}, \textit{r/Anxiety}, \textit{r/ptsd}, \textit{r/depression\_help} are labelled accordingly to the mental health issue they are related to. For subreddits \textit{r/adultsurvivors} and \textit{r/mentalhealth}, the posts are labelled using rule-based methods based on the keywords (\textit{depression}, \textit{panic attack}, \textit{trauma}, etc) used for collection of posts. A manual checkup is given to the posts for making sure that keywords related to a particular mental health issue is not found in posts related to a different mental health issue. Table \ref{tab:table1} ``Traits'' column shows the label for posts collected from different subreddits.\par

\begin{table*}[hbt!]
\centering
\caption{Posts collected from different subreddits}
\label{tab:table1}
\begin{tabular}{|l|l|l|l|l|}
\hline
\textbf{Subreddits} &  \textbf{\# of posts} & \textbf{Final \# of posts} & \textbf{\# of users} & \textbf{Traits} \\ \hline
\textbf{\textit{adultsurvivors}} & 6619 & 6447 & 4523 & mental health issues \\ 
\textbf{\textit{depression}} & 455 & 419 & 400 & depression \\ 
\textbf{\textit{Anxiety}} & 18 & 16 & 13 & anxiety \\ 
\textbf{\textit{ptsd}} & 738  & 714 & 639 & ptsd\\ 
\textbf{\textit{depression\_help}} & 37 & 37 & 37 & depression\\ 
\textbf{\textit{mentalhealth}} & 329 & 324 & 312 & mental health issues \\ \hline
\end{tabular}
\end{table*}

\begin{figure}[hbt!]    
\centering
      \includegraphics[width=0.43\textwidth, height=0.43\textwidth]{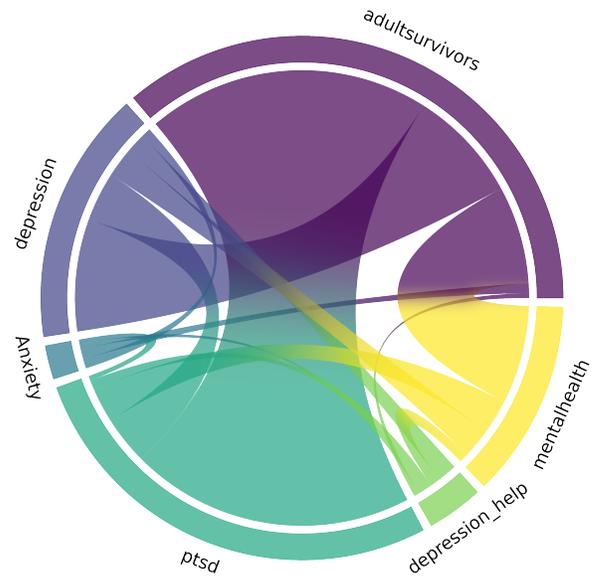}
      \caption{{Intersections of different subreddits}}
        \label{fig:subreddits_network}     
\end{figure} 

\begin{figure}[hbt!]    
\centering
      \includegraphics[width=0.4\textwidth, height=0.34\textwidth]{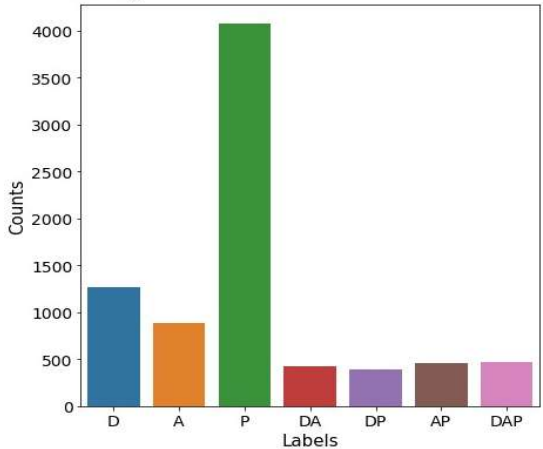}
      \caption{{Distribution of posts}}
        \label{fig:labels}     
\end{figure}

After we collected posts from various subreddits, we looked at intersections (count of common users) (Figure \ref{fig:subreddits_network}) between various mental health subreddits and \textit{r/adultsurvivors} using the author IDs of posts collected. It can be viewed that the arc between \textit{r/adultsurvivors} and \textit{r/ptsd} is the most broad with 58 common users. This depicts the higher chances of CSA survivors having trauma-related issues and it reflects as mentioned in earlier research done by psychologists \cite{aspelmeier2007childhood}. We also notice intersections between different mental health subreddits and it can be interpreted as the possibility of co-occurrence of multiple mental health issues simultaneously. Between mental health subreddits, \textit{r/ptsd} and \textit{r/depression} with 10 common users is the highest. The posts containing multiple mental health issues are labeled accordingly. Figure \ref{fig:labels} presents the distribution of posts among various labels.\par

After labeling the posts we looked at prominent words when different mental health issues occur. Log-odds ratio \cite{towardsdatascienceGenerateMeaningful} is used to score the terms. In comparison to corpora that is related to different mental health issue, the log-odds ratio aids in identifying distinctive prevalent words in a corpus of a certain mental health issue. In the depression (D) corpus, words like fruitlessly and shortcoming have higher scores. The anxiety (A) corpus contains words like fogginess, which are associated with brain fog anxiety. Words connected to memories of the traumatic experience, such as revictimization and surfacing, are highly rated in the PTSD (P) corpus. Prominent words when multiple mental health issues co-occur (DA, AP, DP, DAP) are also provided. Table \ref{tab:table2} and \ref{tab:table3} shows the prominent words when different mental issues occurs.\par

\begin{table*}[hbt!]
\centering
  \caption{Prominent unigrams when different mental health issues occurs. Depression (D), Anxiety (A) and PTSD (P)}
  \label{tab:table2}
  \addtolength{\tabcolsep}{-5.2pt}
  \begin{tabular}{|l|l|l|l|l|l|l|l|l|l|}
    \hline
      \multicolumn{2}{|c|}{\textbf{D}} &
      \multicolumn{2}{c|}{\textbf{A}}  &
      \multicolumn{2}{c|}{\textbf{P}}  &
      \multicolumn{2}{c|}{\textbf{DA}} &
      \multicolumn{2}{c|}{\textbf{DP}}  \\ \hline
    \textbf{Unigram} & \textbf{Score} & \textbf{Unigram} & \textbf{Score} & \textbf{Unigram} & \textbf{Score} & \textbf{Unigram} & \textbf{Score} & \textbf{Unigram} & \textbf{Score}\\\hline
    fruitlessly & 4.75 & fogginess & 5.23 & dominance & 4.43 & paroxetine & 5.72 & entrap & 5.79 \\ 
    shortcoming & 4.26 & foam & 5.23 & devastation & 3.62 & diazepam & 5.72 & dreamscape & 5.79\\
    dissatisfaction & 4.26 & hypochondriacs & 4.82 & revictimization & 3.43 & cyclothymia & 5.30 & insensitivity & 4.79\\
    ephebophilia & 4.26 & emojibator & 4.82 & transference & 3.27 & sponge & 5.30 & dissociation & 4.79\\ 
    exacerbate & 4.26 & submerge & 4.23 & catheter & 3.21 & nervous & 4.72 & leeching & 4.79\\
    absorb & 3.94 & chauvinist & 4.23 & helpless & 3.21 & childtrigger & 4.72 & disheartening & 4.79\\
    suicidally & 3.94 & tainting & 4.23 & cling & 2.94 & colluding & 4.72 & somnophilia & 4.79\\
    subjugation & 3.94 & carelessness & 4.23 & tremble & 2.94 & hypogonadism & 4.72 & blisteringly & 4.79\\
    oxycodone & 3.94 & stuntedness & 4.23 & traumatic & 2.94 & puffing & 4.72 & dysphoric & 4.79 \\
    lamotrigine & 3.94 & vaginal & 4.23 & surfacing & 2.94 & anxiousness & 4.72 & retraumatize & 4.79\\
    \hline
  \end{tabular}
\end{table*}

\begin{table}[hbt!]
\centering
  \caption{Prominent unigrams when different mental health issues occurs. Depression (D), Anxiety (A) and PTSD (P)}
  \label{tab:table3}
  \begin{tabular}{|l|l|l|l|l|l|l|l|l|l|l|l|l|l|l|}
    \hline
      \multicolumn{2}{|c|}{\textbf{AP}} &
      \multicolumn{2}{c|}{\textbf{DAP}} \\ \hline
     \textbf{Unigram} & \textbf{Score} & \textbf{Unigram} & \textbf{Score}\\\hline
     overactivation & 5.41 & incapacity & 5.17\\ 
     nauseating & 5.41 & limerence & 5.17\\
     pistanthrophobia & 5.41 & taunting & 4.76\\
     sensitive  & 4.83 & misperception & 4.76\\ 
     reanswering & 4.82 & rebound & 4.76\\
     propranolol & 4.83 & jeopardize & 4.17\\
     coddling & 4.83 & sodomize & 4.17\\
     reemerge  & 4.83 & sleeplessness & 4.17\\
     hypersexualization & 4.83 & transphobia & 4.17\\
     straining & 4.83 & hypermania & 4.17\\
    \hline
  \end{tabular}
\end{table}




\section{RQ2: Are there differences between the posts related to mental health issues with and without CSA background?} 

\begin{figure}[t!]
    \centering
    \subfloat[with CSA]{{\includegraphics[width=0.38\textwidth,height=0.08\textwidth]{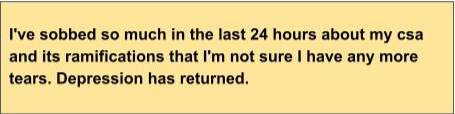}}\label{fig:csa1}}\\
    \subfloat[without CSA]{{\includegraphics[width=0.38\textwidth,height=0.08\textwidth]{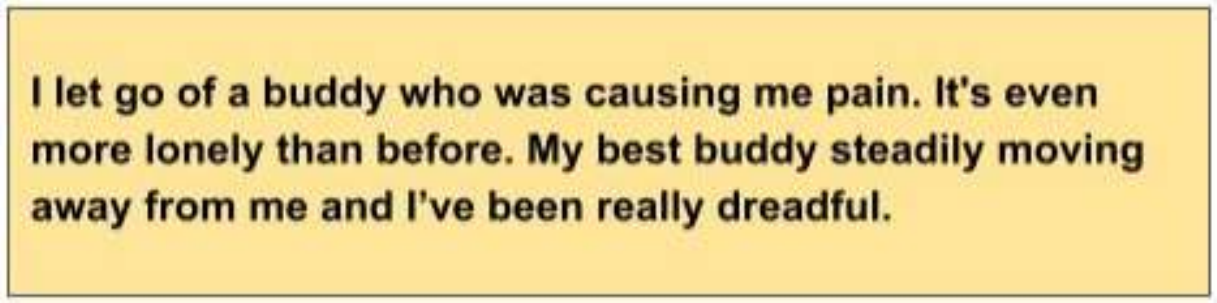}}\label{fig:wcsa1}}
   \caption{Examples of posts with and without CSA background}
\label{fig:egs}
\end{figure}

Mental health issues in users might not be necessarily due to CSA as evident in some of the posts. For example, comparing the posts shown in Figure \ref{fig:egs} with and without CSA background, it is evident that they are pretty similar though small differences may exist. Thus, to understand if posts with CSA background have certain characteristics which are different from posts without CSA background, we explore various NLP methods. Firstly, we need posts without CSA background and for that we used an existing dataset \cite{Kim2020} annotated for various mental health issues. 
The dataset contains posts from various mental health subreddits, so it may have posts with CSA background. Posts that contains words closely related with CSA such as \textit{childhood sexual abuse} and \textit{csa} are removed and finally 9136 posts related to depression, anxiety, PTSD are extracted from the dataset as our work focuses only on these three mental health issues. These posts are again manually inspected to ensure they are not CSA-related.

\label{sub3}
\subsection{Methodology}


We use various NLP approaches such as word-shift, wordcloud, topic analysis, and emotion analysis \cite{neha2022tale}. At first we utilised proportion shift, a type of word-shift \cite{Gallagher2021} graph that ranks words based on scores generated from the difference between a word's relative frequency in first textual content and its relative frequency in the second textual content. For both posts with and without CSA background, we intended to look at unique unigrams and bigrams. To score unigrams and bigrams, TF-IDF (Term frequency-Inverse document frequency) is used and the results are displayed as a wordcloud. Topic analysis is also done for discovering various topics in both posts with and without CSA background using BERTopic \cite{grootendorst2022bertopic} and words inside each topic are scored using c-TF-IDF (class-based TF-IDF). Lastly, we use fine-tuned DistilRoBERTa \cite{hartmann2022emotionenglish} approach for detecting emotions for both posts with and without CSA background. This approach labels the posts to the most probable emotion among the emotions, namely, anger, sadness, fear, disgust, neutral, surprise, and joy.\par  


\subsection{Results} 
\label{subs3}

\begin{figure}[hbt!]    
\centering
      \includegraphics[width=0.3\textwidth, height=0.46\textwidth]{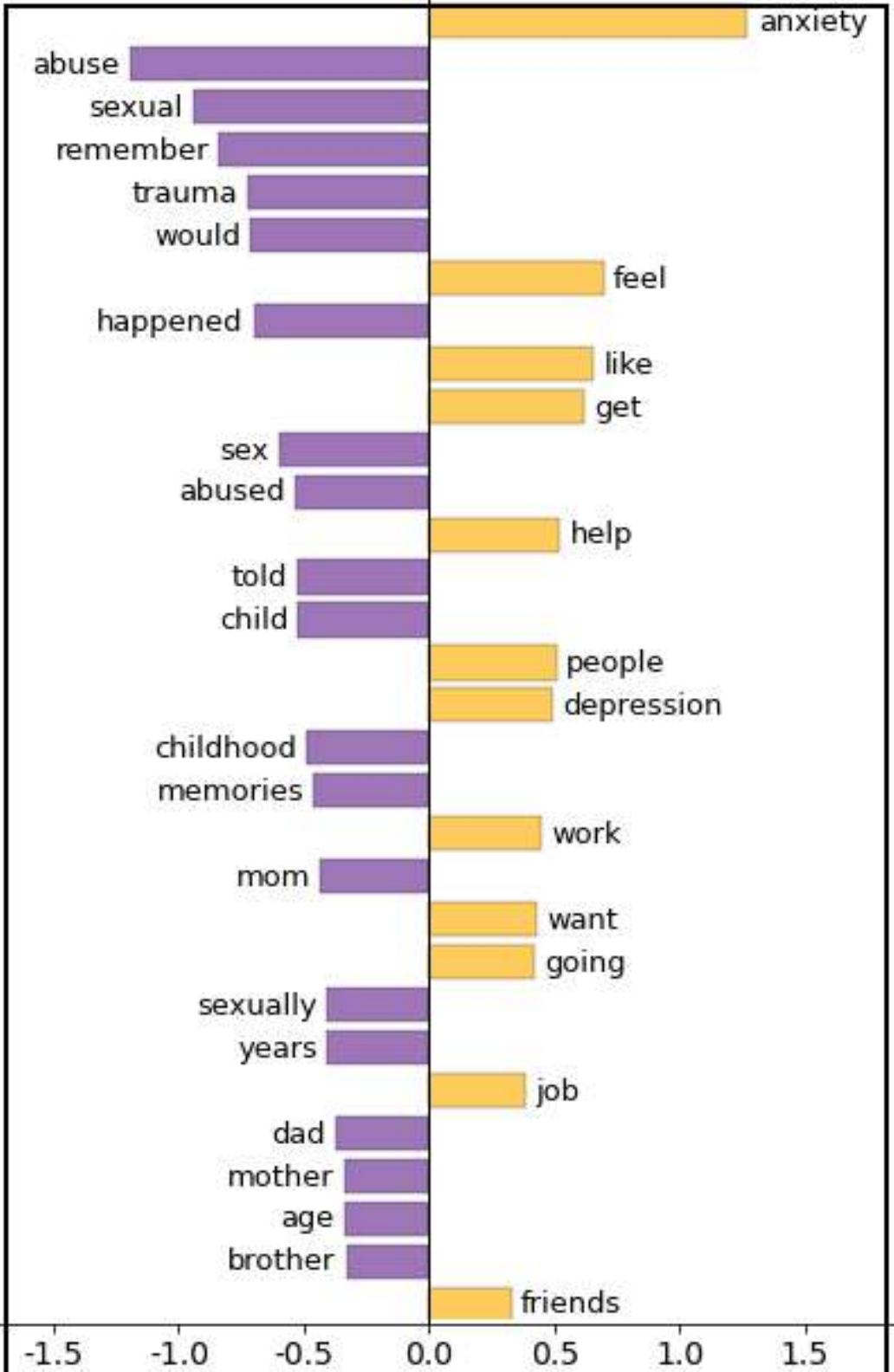}
      \caption{Word shift for posts with CSA background \textit{(left)} and without CSA background \textit{(right)}}
        \label{fig:proportion_shift}     
\end{figure}

As visualized in Figure \ref{fig:proportion_shift}, words that describe CSA (``abuse'', ``sexual'', and ``child'') are prominent in posts with CSA background and this helps to paint a clear picture of how CSA could be the core cause of mental health issues in CSA survivors. In contrast to posts without CSA background, words that represent memory or flashbacks of the traumatic event (``remember'', ``happened'', ``memories'') are prominent in posts with CSA background. From these words, it can be deduced that memories or flashbacks of the traumatic event can play a vital role in the development of mental health issues such as PTSD (``trauma'') in CSA survivors (Section \ref{sub2}). In contrast to posts without CSA background, words that demonstrate familial relationships (``dad'', ``brother'', and ``mom'') stand out in posts with CSA background which may point out the likelihood of perpetrators being family members \cite{Finkelhor1994-qn}. For example, phrases such as ``..cried..over..\textbf{csa}..'', ``..\textbf{memory}..being..raped..'', ``..\textbf{remembered}..stuff..'' appeared often in the corpus with CSA background.\par

Words associated with the workplace (``work'' and ``job'') are more prominent in posts without CSA background than in posts with CSA background, indicating that the development of mental health difficulties may be linked to the workplace. Words ``anxiety'' and ``depression'' 
are more common in posts without CSA background, indicating that these mental health concerns are more prevalent. For example, without CSA background,
``..\textbf{depression} comes back..'', ``..\textbf{depression} suicidal thoughts..f**king \textbf{feeling}..'', ``need..\textbf{friend}..''.\par

\begin{figure}[t!]
    \centering
    \subfloat[with CSA]{{\includegraphics[width=0.25\textwidth]{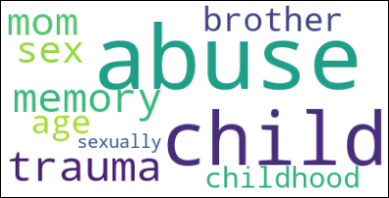}}\label{fig:unicsa}}
    \subfloat[without CSA]{{\includegraphics[width=0.25\textwidth]{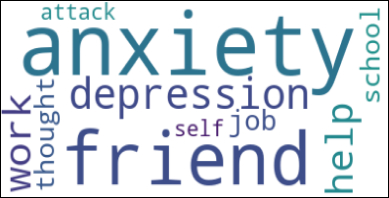}}\label{fig:uniwcsa}}
   \caption{Wordcloud for top unigrams of posts with and without CSA background}
\label{fig:Uni}
\end{figure}

\begin{figure}[t!]
    \centering
    \subfloat[with CSA]{{\includegraphics[width=0.25\textwidth]{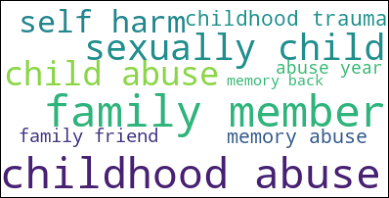}}\label{fig:bicsa}}
    \subfloat[without CSA]{{\includegraphics[width=0.25\textwidth]{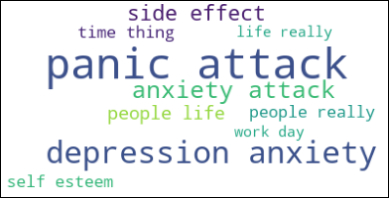}}\label{fig:biwcsa}}
   \caption{Wordcloud for top bigrams of posts with and without CSA background}
\label{fig:bi}
\end{figure}

\begin{table}[hbt!]
\centering
  \caption{Top topics and its top five N-grams for posts with CSA background}
  \label{tab:table5}
  \begin{tabular}{|l|l|l|l|}
    \hline
          \multicolumn{2}{|c|}{\textbf{Topic 1}} &
      \multicolumn{2}{c|}{\textbf{Topic 2}} \\ \hline
      \textbf{N-gram} & \textbf{Score} & \textbf{N-gram} & \textbf{Score}\\\hline
      abuse & 0.0183 & doctor & 0.0924\\
      trauma & 0.0126 & exam & 0.0545\\
      family & 0.0124 & appointment &  0.0256\\
      sex & 0.0111 & pain & 0.0156\\
      parent & 0.0103 & advice & 0.0119 \\
      \hline
  \end{tabular}
\end{table}

\begin{table}[hbt!]
\centering
  \caption{Top topics and its top five N-grams for posts without CSA background}
  \label{tab:table6}
      \begin{tabular}{|l|l|l|l|}
       \hline
      \multicolumn{2}{|c|}{\textbf{Topic 1}} &
      \multicolumn{2}{c|}{\textbf{Topic 2}} \\ \hline
      \textbf{N-gram} & \textbf{Score} & \textbf{N-gram} & \textbf{Score}\\\hline
      anxiety & 0.0096 & bed & 0.0319 \\
      depression & 0.0067 & anxiety  & 0.0147\\
      work & 0.0067 & week sleep & 0.0078 \\
      school & 0.0062 & trouble & 0.0069\\
      problem & 0.0047 & time sleep & 0.0069\\\hline
      \multicolumn{2}{|c|}{\textbf{Topic 3}} &
      \multicolumn{2}{c|}{\textbf{Topic 4}} \\ \hline
      \textbf{N-gram} & \textbf{Score} & \textbf{N-gram} & \textbf{Score}\\\hline
      calorie & 0.0207  & nausea & 0.0502 \\
      weight loss & 0.0144 & symptom  & 0.0159\\
      fat & 0.0116 & vomit & 0.0150 \\
      exercise & 0.0096 & lack appetite & 0.0083\\
      anorexia & 0.0094 & nausea anxiety & 0.0079\\\hline
      \multicolumn{2}{|c|}{\textbf{Topic 3}} &
      \multicolumn{2}{c|}{\textbf{Topic 4}} \\ \hline
      \textbf{N-gram} & \textbf{Score} & \textbf{N-gram} & \textbf{Score}\\\hline
      ugly & 0.0185 & weed & 0.0621\\
      insecure & 0.0148 & cbd & 0.0571\\
      appearance & 0.0140 & cbd oil & 0.0253\\
      body dysmorphia & 0.0103 & anxiety depression & 0.0130\\
      self image & 0.0097 & weed yesterday & 0.0111\\\hline
      \end{tabular} 
      
\end{table}

In addition to word-shift, we also created wordcloud of the top ten unigrams and bigrams. Figure \ref{fig:Uni} presents the top unique unigrams and Figure \ref{fig:bi} presents the top unique bigrams for posts with and without CSA background. These results converges with those shown in Figure \ref{fig:proportion_shift} but the method used to score the unigrams and bigrams is different than the method used in word-shift. Additionally, bigram such as ``self harm'' related to suicidal intent is more prevalent in posts with CSA background.

Next, to understand various topics which are discussed in both with and without CSA background posts, we perform topic analysis. Two topics are identified using BERTopic in posts with CSA background based on coherence score. Topic 1 is associated with CSA, such as cause (``abuse'') and its after-effect (``trauma''). Topic 2 is concerned with clinical aspects (``doctor'', ``appointment''). Table \ref{tab:table5} displays the topics and top five N-grams for posts with CSA background. In posts without CSA background, thirty topics are discovered based on coherence score, but we only provide the top six topics from the thirty topics obtained and their top five N-grams in Table \ref{tab:table6}. Various topics are discovered including mental health concerns relating to the workplace (``anxiety'', ``depression'', ``work''), eating disorders (``anorexia''), and substance use (``cbd'', ``cbd oil'', ``weed yesterday''). 

\begin{figure}[hbt!]    
\centering
      \includegraphics[width=0.3\textwidth, height=0.3\textwidth]{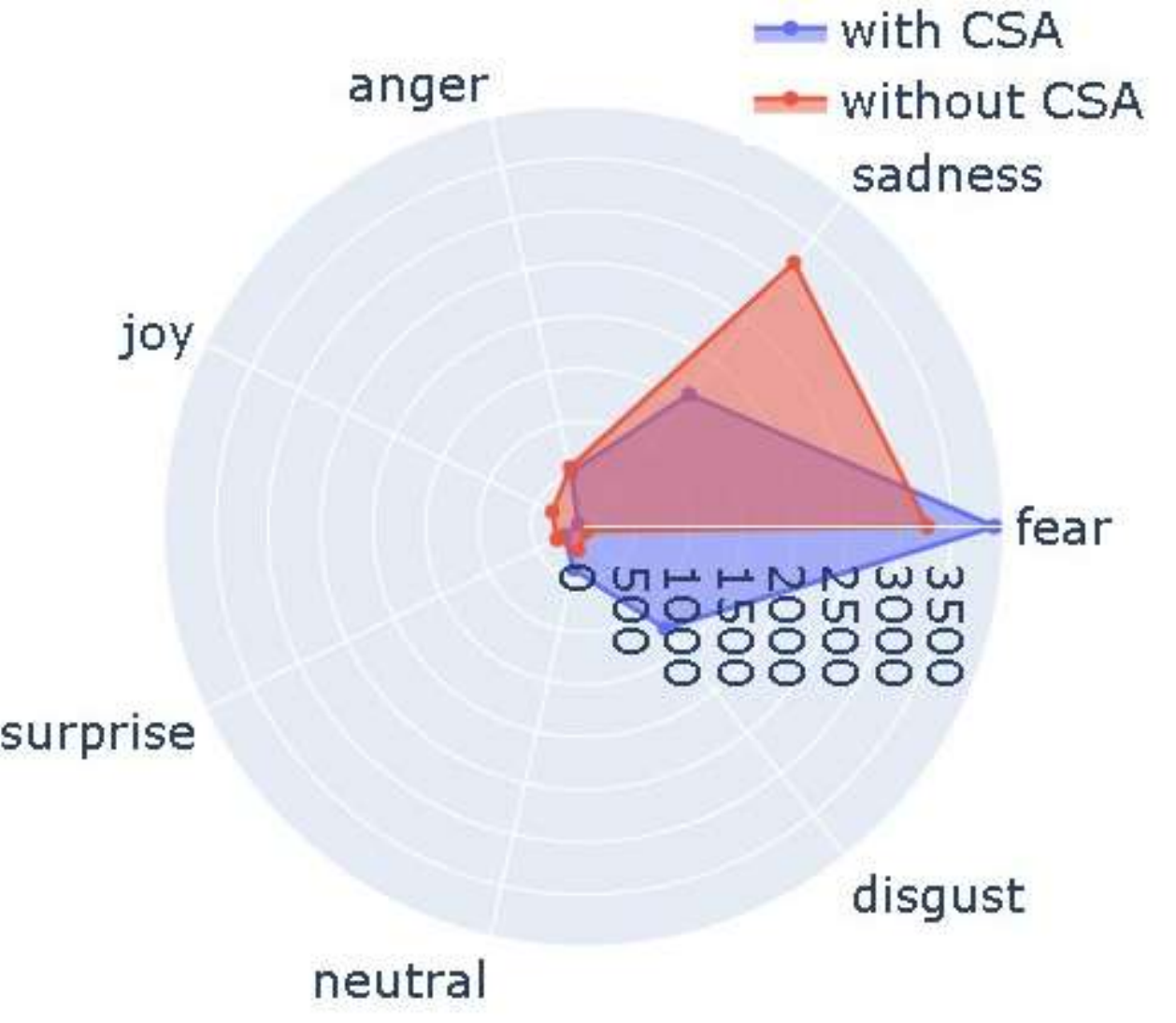}
      \caption{{Emotions of posts related to mental health issues with and without CSA background}}
        \label{fig:emo_analysis}     
\end{figure}

Finally, we also examined if the emotions being present in posts with CSA background are different from those without CSA background. Figure \ref{fig:emo_analysis} shows the emotions for posts with and without CSA background. Emotion ``disgust'' and ``fear'' are associated more with posts with CSA background than posts without CSA background as strong correlation is there between these two emotions and trauma-related issues \cite{coyle2014emotions} and the presence of trauma-related issues is more in posts with CSA background. ``sadness'' emotion is reported more in posts without CSA background as ``sadness'' is associated with depression and it is one of the most common mental health issue. 

\section{RQ3: Can fine-tuning domain-specific transformer model help in identifying mental health issues in CSA survivors from the OSM posts?}

For identifying mental health issues in posts related to CSA, differentiating posts with and without CSA background is crucial as sometimes subtle differences exist between posts with and without CSA background as reported in Section \ref{sub3}. To answer it, we develop a two-stage framework that helps in differentiating posts with and without CSA background and also identification of mental health issues in posts with CSA background. \par 
\label{sub4}

\subsection{Methodology}
\label{subsfm}
\textbf{Proposed Framework:} The first stage entails binary classification of mental health issues related posts into ``with CSA'' and ``without CSA'' background. Along with employing classical machine learning algorithms such as Naive Bayes, Random Forest, and XGBoost, we also use transformer-based models BERT and MentalBERT. We add fully connected layers on top of BERT and MentalBERT \cite{ji2021mentalbert} model's layers and fine-tune them by retraining all the weights of the model's layers on our training data. We design the second stage as multi-label classification for identifying mental health issues in posts with CSA background as multiple mental health issues can occur simultaneously (Section \ref{sub2}). We use tree-based classical machine learning algorithms (XGBoost, Random Forest), Multilayer perceptron (MLP), transformer-based models as used in the first stage. In this stage, we fine-tune BERT and MentalBERT only on training set of posts with CSA background. We set each label's threshold probability as 0.5.

\noindent \textbf{Data split:} For the first stage, we combined both posts with CSA (7957 posts) and without CSA (9136 posts) background. Then we performed a split of 80:20 (13674 and 3419 posts). Train and validation sets contains 12306 and 1368 posts, respectively, out of 13674 posts. Test set consists of 3419 posts. Only posts with CSA background are considered for the second stage, with an 80:20 split (6365 and 1592 posts). The train and validation sets each have 5728 and 637 posts out of 6365 posts. There are 1592 posts in the test set. \par

\noindent \textbf{Model Training:} BERT and MentalBERT are trained on posts with and without CSA background for 10 epochs with a batch size of 8 and a learning rate of 5e-5 for the first stage. For multi-label classification, the number of epochs and learning rate is kept the same for training BERT and MentalBERT as the first stage. Also, text data is converted into embedding vectors before providing as input to classical machine learning models.\par

\noindent \textbf{Model interpretation: }When dealing with a sensitive domain like psychopathology interpreting the model's decision is critical. Integrated gradients (IG) \cite{sundararajan2017axiomatic} method is used to interpret the model's decision. Both the stages of the framework are accompanied by interpretation. IG assigns attribution values to input features, which aids in understanding why the model makes a particular decision. \textit{Alibi} \cite{JMLR:v22:21-0017} library is used to implement integrated gradients. \par

\begin{table}[hbt!]
\centering
\caption{Evaluation results of models for classifying posts to ``with CSA'' and ``without CSA'' background}
\label{tab:table8}
\begin{tabular}{lll}
\hline
\textbf{Model} & \textbf{Accuracy} & \textbf{F1-score (macro)} \\ \hline
\textbf{Naive Bayes} & 0.5607 & 0.3926\\
\textbf{Random Forest} & 0.6774 & 0.6753\\
\textbf{XGBoost} & 0.7452  & 0.7429\\ 
\textbf{BERT} & \textbf{0.9626} & \textbf{0.9624}\\ 
\textbf{MentalBERT} & 0.9509 & 0.9508\\ \hline
\end{tabular}
\end{table}

\begin{table}[hbt!]
\centering
\caption{Evaluation results of models for multi-label classification of mental health issues}
\label{tab:table9}
\begin{tabular}{ll}
\hline
\textbf{Model} & \textbf{Hamming score}\\\hline
\textbf{MLP} & 0.4102\\
\textbf{XGBoost} & 0.5636\\
\textbf{Random Forest} & 0.5698\\ 
\textbf{BERT} & 0.5798\\ 
\textbf{MentalBERT} & \textbf{0.6709}\\ \hline
\end{tabular}
\end{table}

\subsection{Results}

In binary classification, among classical machine learning algorithms XGBoost outperformed both Naive Bayes and Random Forest. Fine-tuned BERT, on the other hand, surpassed all the methods, with an accuracy of 96.26\% and f1-score (macro) of 96.24\%. This shows using a pretrained model and fine-tuning on our dataset lead to improved performance. The models and their scores are shown in Table \ref{tab:table8}. Multi-label classification models are assessed using hamming score. With a hamming score of 67.09\% fine-tuned MentalBERT surpassed other models for multi-label classification. This demonstrates that model pre-trained on mental health data is more effective at identification of mental health issues in CSA survivors. Table \ref{tab:table9} illustrates the models and their corresponding scores.\par


\subsection{Visualizing interpretation of model's response}

\begin{figure}[t!]
    \centering
    \subfloat[with CSA]{{\includegraphics[width=0.44\textwidth, height = 0.028\textwidth]{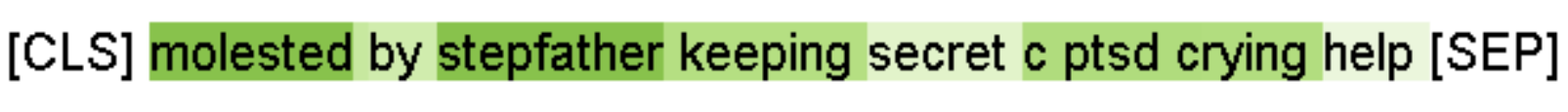}}\label{fig:excsa}}\\
    \subfloat[without CSA]{{\includegraphics[width=0.435\textwidth, height = 0.05\textwidth]{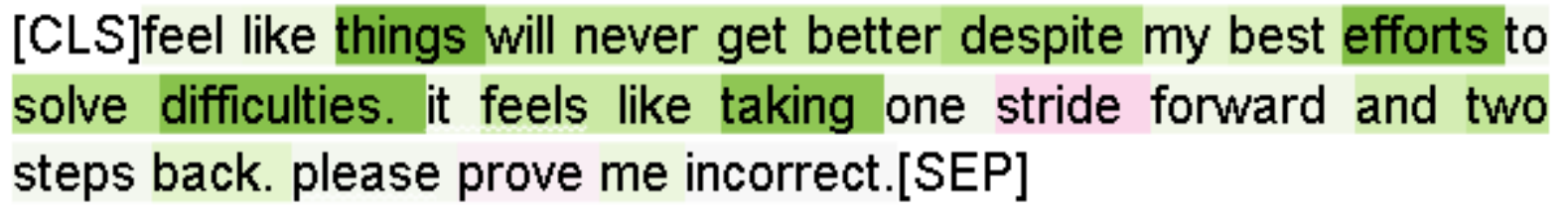}}\label{fig:exwcsa}}
   \caption{Interpretation of fine-tuned BERT model's decision on true positive (top) and true negative (bottom) instances}
\label{fig:exbi}
\end{figure}

\begin{figure}[t!]    
\centering
      \includegraphics[width=0.41\textwidth, height = 0.06\textwidth]{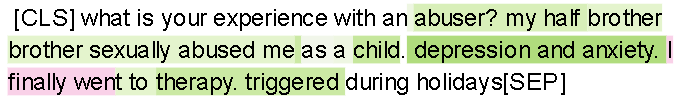}
      \caption{{Interpretation of fine-tuned MentalBERT model's decision on true positive instance }}
        \label{fig:exml}     
\end{figure}

For both stages, we visualize the interpretations of the best model's decision. Figure \ref{fig:exbi} shows the explanations on two instances for fine-tuned BERT. Figure \ref{fig:excsa} shows the true positive (correctly classified as ``with CSA'' background) and words that are highlighted in green contributed for this decision and pink represents those that contributed oppositely towards this decision. Phrases such as ``molested by'', ``stepfather keeping'', ``ptsd crying'' are highlighted in green which explains that fine-tuned BERT is able to correctly identify phrases that denote the traumatic event, perpetrator, and trauma related issues commonly related to posts with CSA background. Figure \ref{fig:exwcsa} shows the true negative (correctly classified as ``without CSA'' background). Phrases ``things like never get better'', ``over difficulties'' are highlighted in green which could relate to sadness but are not related to CSA. Figure \ref{fig:exml} shows the explanations for fine-tuned MentalBERT on a true positive instance (correctly labeled as depression and anxiety). Phrases such as ``depression and anxiety set in'' are highlighted in green which shows that fine-tuned MentalBERT is able to capture information that is related to depression and anxiety. The phrase ``usually triggered'' that pertains to anxiety triggers is also highlighted in green. 

\section{Conclusion}

As far as we are aware, our work is a pioneer for exploring OSM for the analysis of mental health issues in CSA survivors as previous studies have not paid particular attention towards gaining a better understanding of mental health issues among CSA survivors. We discovered the commonly discussed mental health issues in posts with CSA background and for achieving this, we created a dataset by collecting mental health issues related posts with CSA background from multiple subreddits. Our research also found that minor variations exist between posts with and without CSA background. We developed a two-stage framework for identifying mental health issues in CSA survivors, with the first stage identifying posts with CSA background while taking into consideration the differences between posts with and without CSA background and followed by multi-label classification of mental health issues with posts reported to be related with CSA in second stage. The best model in first stage reported highest accuracy and f1-score (macro) of 96.26\% and 96.24\% respectively and in second stage, the best model reported highest hamming score of 67.09\%. We have multiple plans to extend this work. In our future work, we would like to include more subreddits for our analysis. One of the reasons of including more subreddits is to include additional mental health issues in our analysis.



\bibliographystyle{IEEEtran}
\bibliography{main_IEEE}

\end{document}